\documentclass[twocolumn,showpacs,prl,amsmath,amssymb]{revtex4}

 \usepackage{graphicx}
 \usepackage{dcolumn}
 \usepackage{bm}

\begin{document}
 \title{Homotopy Theory of Topological Defects in Spinor Condensates}
 \author{Yunbo Zhang$^{1,2}$, Harri M\"{a}kel\"{a}$^{1}$, and 
 Kalle-Antti Suominen$^{1,3}$}

 \affiliation{$^{1}$Department of Physics, University of Turku, 
 FIN-20014 Turun yliopisto, Finland\\
 $^{2}$Department of Physics and Institute of Theoretical Physics, 
 Shanxi University, Taiyuan 03006, P. R. China\\
 $^{3}$Helsinki Institute of Physics, PL64, FIN-00014 Helsingin 
 yliopisto, Finland}
 \date{\today}
 \begin{abstract}
 We investigate the topological defects in atomic spin-1 and spin-2 
 Bose-Einstein condensates by applying the homotopy group theory. With 
 this rigorous approach we clarify the previously controversial 
 identification of symmetry groups and order parameter spaces for the 
 spin-1 case, and show that the spin-2 case provides a rare example of 
 a physical system with non-Abelian line defects, and the possibility 
 to have winding numbers of $1/3$ and its multiples.
 \end{abstract}

 \maketitle

 The all-optical trapping of Bose-Einstein condensates (BEC) 
 \cite{Optical} has opened up a
 new direction in the study of dilute atomic gases, i.e., the spinor 
 condensates with degenerate internal degrees of freedom of the 
 hyperfine spin $F$. For alkali atoms with $F=1$, both experiments and 
 theories have shown two possible kinds of spin correlations in the 
 atom species, namely ferromagnetic (e.g. $^{87}$Rb) or 
 antiferromagnetic (e.g. $^{23}$Na). With the experimental success of 
 condensing alkali bosons with $F>1$ such as $^{85}$Rb and 
 $^{133}$Cs~\cite{CornishWeber}, and the unusual stability of the 
 $F=2$ state (against spin-exchange) in $^{87}$Rb~\cite{Suominen}, one 
 expects that defects with much richer structure can be created in the 
 future. A remarkable feature here is that not only the
 gauge symmetry $\rm{U(1)}$ but also the spin symmetry $\rm{SO(3)}$ are 
 involved, a situation similar to superfluid $^{3}$He where three 
 different continuous symmetries (orbital, spin and gauge) are broken 
 either independently or in a connected fashion~\cite{Helium}.

 Topological defects and excitations in the spinor BECs have been 
 studied theoretically
 by several groups \cite{Ho,Ohmi,Law,Stoof,Khawaja,Zhou1,Martikainen}. 
 Most of the work on this subject is based on the original 
 identifications of the order parameter spaces by Ho~\cite{Ho}. After 
 the original studies it was also claimed by Zhou that a discrete 
 symmetry of $Z_{2}$ type was missed in the case of antiferromagnetic 
 spin-1 condensate~\cite{Zhou1} and therefore the topological defects 
 would manifest totally different structures. In this Letter we 
 present a rigorous topological study that both solves this 
 spin-1 controversy, and reveals interesting aspects of spin-2 
 systems. The phases of spin-2 spinor condensates are characterized by 
 a pair of order parameters $\left\langle {\bf F}\right\rangle $ and 
 $| \Theta |$ describing the ferromagnetic 
 order and the formation of singlet pairs, respectively 
 \cite{Ciobanu,Koashi,MartikainenJPB}. It turns out that for the so 
 called cyclic phase the fundamental group that determines the nature 
 of possible stable topological defects is {\it non-Abelian}. The only 
 known physical example of such a system so far has been the biaxial 
 nematic liquid crystal.

 {\em Outline of the homotopy theory of defects}: We sketch out the procedure
 which has been widely used in the study of topological defects in ordered
 media such as liquid crystals, superfluid $^{3}$He and heavy-fermion
 superconductors \cite{Mermin}. The central feature of the classification
 scheme of the defects emerges from examining the mappings of closed curves
 in physical space into the order-parameter space (OPS). The order parameter
 of a spinor BEC, $\Psi ({\bf r})=\sqrt{n({\bf r})}\zeta ({\bf r})$,
 where $n({\bf r})$ is the density and $\zeta ({\bf r})$ is a normalized
 spinor $\zeta ^{\dagger }({\bf r})\cdot \zeta ({\bf r})=1$, has associated
 with it a group of transformations $G$, i.e., all spinors related to each
 other by gauge transformation $e^{i\theta }$ and spin rotations
 $e^{-iF_{z}\alpha }e^{-iF_{y}\beta }e^{-iF_{z}\gamma }$ are energetically 
 degenerate in zero external magnetic field, where $(\alpha ,\beta ,\gamma )$ are the Euler
 angles. The set of all transformations in $G$ that leave the reference order
 parameter (chosen arbitrarily but thereafter fixed) unchanged is known as the
 isotropy group $H$. The OPS can then be taken to be the space of 
 cosets of $H$ in $G$: $M=G/H$. In terms of broken symmetry, the fact 
 that the ordering
 breaks the underlying symmetry is expressed in the fact that $H$ is only a
 subgroup of $G$. Let $G$ be a {\it connected}, {\it simply connected} 
 continuous group
 and $H_{0}$ be the set of points in $H$ that can be connected to the identity
 by a continuous path lying entirely in $H$. The fundamental theorems assert that
 \begin{equation}
     \pi _{1}(M)=H/H_{0}, \qquad \pi _{2}(M)=\pi _{1}(H_{0}).
 \end{equation}
 where the fundamental (second) homotopy group $\pi_{1} (\pi_2)$ 
 characterizes the
 singular line (point) defects in three-dimensional space.

 For the spinor condensate it seems natural to identify the underlying
 symmetry group as ${\rm U(1) \times SO(3)}$, the groups in the direct product
 representing the gauge and spin degrees of freedom respectively. This
 group is {\it not} simply connected, i.e., $\pi_1({\rm U(1) \times SO(3)}) \ne 0$.
 To apply the theorems, however, it is essential that one chooses the group $G$
 to be simply connected (i.e., $\pi_1(G)=0$). This can always be done 
 because any continuous group can be imbedded in its universal 
 covering group. We proceed by specifying the symmetry group as 
 $R\times{\rm SU(2})$, with the group of real numbers $R$ representing 
 any translation $\theta \in (-\infty,+\infty )$ in the phase of the condensate.
 For $F=1$, we use the 3D representation of  the group ${\rm SU(2)}$ in order to 
 obtain the isotropy group, e.g., a  rotation $u(z,\alpha)$ around axis $z$ 
 by angle $\alpha $ takes the form of
 a diagonal matrix $Diag\left( e^{-i\alpha },1,e^{i\alpha }\right) $.
 The element $-u(z,\alpha)$ is represented by the same matrix, i.e.,
 $u(z,\alpha +2\pi )=-u(z,\alpha )$ while $u(z,\alpha +4\pi )=u(z,\alpha )$.

 {\em Calculation of the homotopy groups: }There are two possible
 ground states in $F=1$ case. For the ferromagnetic state, the
 isotropy group $H$ is constructed by the set of transformations which
 leave the reference order parameter $(1,0,0)^{T}$ invariant. From the
 ground state spinor $\zeta =e^{i\left( \theta -\gamma \right) }\left(
 e^{-i\alpha }\cos ^{2}\frac{\beta }{2},\sqrt{2}\cos \frac{\beta }{2}\sin
 \frac{\beta }{2},e^{i\alpha }\sin ^{2}\frac{\beta }{2}\right) ^{T}$ we know
 immediately that the angles should satisfy
 \begin{equation}
     \beta =0, \theta -\alpha -\gamma =2n\pi
 \end{equation}
 with $n$ an integer. The elements in group $H$ are the combination of a
 translational part and a rotational part $H=\left\{ \left( \theta,
 u(z,\theta )\right) ,\left(\theta,u(z,\theta +2\pi )\right) \right\}
 =\left\{ \left( \theta,\pm u(z,\theta )\right) \right\} $. Evidently
 this group includes two disconnected components---the connected component of
 the identity $H_{0}=\left\{ \left( \theta,u(z,\theta )\right) \right\} $ is isomorphic
 to $R$. The group $H/H_{0}$ is isomorphic to the
 integers modulo 2, i.e., $Z_{2}$. The second homotopy group $\pi _{2}$ is
 trivial and we arrive at the same result as that in Ref. \cite{Khawaja}
 \begin{equation}
     \pi _{1}(M)=Z_{2},\qquad \pi _{2}(M)=0\text{, (spin-1 FM state). }
 \end{equation}
 A ferromagnetic spin-1 condensate may have therefore only singular vortices
 with winding number one while the point-like defects are 
 topologically unstable.

 The polar state emerges if the atoms in the condensate interact
 anti-ferromagnetically. In the ground state $\zeta =e^{i\theta }/\sqrt{2}
 \left( -e^{-i\alpha }\sin \beta ,\sqrt{2}\cos \beta ,e^{i\alpha }\sin \beta
 \right) ^{T}$, the reference parameter $(0,1,0)^{T}$ is left invariant for
 just those elements with
 \begin{equation}
     \beta =0,\theta =2n\pi \quad \text{ or }\quad \beta =\pi ,\theta 
 =\left( 2n+1\right) \pi.
 \label{p1}
 \end{equation}
 Thus the isotropy group $H$ must now be expanded to include not only
 transformations in which both the rotation and the translation leave 
 the spinor unchanged, but also those in which
 the rotation takes the reference spinor $(0,1,0)^{T}$ to 
 $(0,-1,0)^{T}$ and the translation
 takes it back, i.e., a $\pi $ rotation about arbitrary axis perpendicular to
 $\hat{z}$ combined with a $\pi $ translation in $\theta $ (or any odd
 multiples of $\pi $). The full isotropy group is the union of these two sets,
 $H=\left\{ \left( 2n\pi,u(z,\alpha )\right), \left( (2n+1)\pi,gu(z,\alpha )\right) \right\} $ 
 where $g=u(y,\pi)$. There are infinitely many discrete 
 components in $H$, while the connected component of the identity 
 $H_{0}=\left\{ \left( 0,u(z,\alpha)\right) \right\}$ is isomorphic to $\rm{U(1)}$. 
 The elements with an even translational parity are of the form 
 $(2n\pi,I)H_{0}$, and those with an odd parity are of the form 
 $\left( (2n+1)\pi,g\right) H_{0}$. The group $H/H_{0}$ is therefore
 isomorphic to the group of integers $Z$ through the isomorphism
 $\left( (2n+j)\pi,g^{j}\right)H_0 \mapsto 2n+j$ for $j=0,1$.
 We recover the conclusion that line and
 point defects in spin-1 polar state can be classified by integer winding
 numbers,
 \begin{equation}
     \pi _{1}(M)=Z,\qquad \pi _{2}(M)=Z\text{, (spin-1 Polar state).}
 \end{equation}
 Thus the $Z_{2}$ term does not appear in the homotopy
 group. As we shall see the identification of the OPS in 
 Ref.~\cite{Zhou1} is also incorrect.

 {\em Spin-2 Bose condensate}: We next apply the same approach to the BEC of
 spin-2 bosons. There are three possible phases for a spin-2
 spinor Bose condensate, i.e., one more compared to the spin-1 case. This extra
 phase comes from the additional interaction parameter describing the 
 singlet formation, and is referred to as the {\em cyclic} 
 phase~\cite{Ciobanu,Koashi,MartikainenJPB}. The defects which may be created 
 in spin-2 condensate
 exhibit even more elaborate structures due to quantum correlations among
 bosons. For $F=2$ we have to use the 5D representation of $SU(2)$, e.g., the
 rotation $u(z,\alpha )$ is represented by matrix $Diag\left( e^{-2i\alpha
 },e^{-i\alpha },1,e^{i\alpha },e^{2i\alpha }\right) $. The calculations of
 the degenerate family of the ground state spinors and the corresponding
 homotopy groups are straightforward and details will be presented elsewhere
 \cite{Harri}. Here we only pick up some interesting features in our results,
 focusing on the symmetry properties of the defects in comparison with those
 in other ordered media.

 Equating the general expression for the ground state spinor of the
 ferromagnetic state $F$
 \begin{equation}
     \zeta =e^{-i(\theta -2\gamma )}\left(
     \begin{array}{c}
        e^{-2i\alpha }\allowbreak \cos ^{4}\frac{\beta }{2} \\
        e^{-i\alpha }\sin \beta \cos ^{2}\frac{\beta }{2} \\
        \frac{\sqrt{6}}{4}\sin ^{2}\beta \\
        e^{i\alpha }\sin \beta \sin ^{2}\frac{\beta }{2} \\
        e^{2i\alpha }\sin ^{4}\frac{\beta }{2}
     \end{array}
     \right)  \label{f}
 \end{equation}
 with the reference spinor $(1,0,0,0,0)^{T}$ leads to the requirement for the
 isotropy group $H$
 \begin{equation}
     \beta =0,\theta -2\alpha -2\gamma =2n\pi .
 \end{equation}
 We see that taking $n=0,1,2,3$ is enough for all possible transformations,
 with the translational part being arbitrary and the rotational part
 containing the rotations around axis $z$ by $\theta/2,\theta/2+\pi ,
 \theta/2+2\pi ,\theta/2+3\pi $ respectively.
 Hence the group $H$ is composed of 4 pieces $H=\left\{ \left( \theta,
 u(z,n\pi +\theta /2)\right) \right\} $. Here it is important to show that
 the 4 components are not connected: there does not exist a continuous path in 
 $H$ which connects one component to another, though the rotational parts
 themselves are connected. The connected component of the identity $
 H_{0}=\left\{\left(\theta,u(z,\theta /2)\right) \right\}$ is again isomorphic to 
 $R$. If we define an element $g$ of the group $R\times \rm{SU(2)}$ by
 $\left(0,u(z,\pi )\right) $, we see that the quotient group $H/H_{0}$ has the
 same structure as the cyclic group of order 4, i.e., $\{e,g,g^{2},g^{3}\}$
 and we conclude that
 \begin{equation}
     \pi _{1}(M)=Z_{4},\qquad \pi _{2}(M)=0\text{, (spin-2 }F\text{ state).}
     \label{fm2}
 \end{equation}

 It is interesting to check how the group $Z_{4}$ characterizes vortices for
 state $F$. In spin-1 case there is only one topologically stable line defect, 
 that is, a vortex with winding number one. Equation (\ref{fm2}) shows that 
 there are 3 stable vortices for spin-2 condensates. We can set 
 $\theta -2\gamma =2m\varphi$, $-\alpha =m\varphi$, $\beta =\pi t$ 
 in the ground state for $F$ state, Eq. (\ref{f}), which
 leads to a family of spinor states parametrized by a parameter $t$ between 0
 and 1. Here $m>0$ is an integer, $\varphi $ is the azimuthal angle. When $t$
 evolves from 0 to 1, the $4m\varphi $ vortex state $\zeta (t=0)=\left(
 e^{i4m\varphi },0,0,0,0\right) ^{T}$ evolves continuously to the vortex free
 state $\zeta (t=1)=\left( 0,0,0,0,1\right) ^{T}$. This shows that
 vortices with winding number $4m$ are topologically unstable. Similarly, 
 by multiplying factors $
 e^{ik\varphi }(k=1,2,3)$ one obtains the following correspondences
 \begin{equation}
     e^{i\left( 4m+k\right) \varphi }\left( 1,0,0,0,0\right) ^{T}\rightarrow
     e^{ik\varphi }\left( 0,0,0,0,1\right) ^{T}
 \end{equation}
 i.e., the vortices with winding numbers $4m+k$ may evolve into vortices with
 winding numbers $k$, respectively. There are thus 3 classes of topologically stable
 line defects. Together with the uniform state, they form the 
 fundamental group $Z_{4}$.
 Straightforwardly for condensates with spin $F,$ the fundamental group $\pi
 _{1}(M)=Z_{2F}$ characterizes $(2F-1)$ classes of stable line defects.

 {\em Non-Abelian homotopy groups}: Media with non-Abelian fundamental groups
 are especially interesting from the topological point of view. The only
 illustrative example in ordered media so far have been biaxial nematic liquid
 crystals~\cite{Yu}. If $G$ is
 taken to be $\rm{SO(3)}$ then the isotropy group $H$ is the four-element group, $
 D_{2}$, the symmetry group of a rectangular box. If, however, we take $G$ to
 be $\rm{SU(2)}$, then $H$ is expanded to the non-Abelian quaternion group $Q$
 (known as the {\it lift} or {\it double group}) with eight elements, 
 the multiplication table of which has been verified experimentally
 \cite{DeNeve}.

 We have found that the cyclic state $C$ provides another physically realistic
 example in which the fundamental group is non-commutative. The reference spinor
 $\frac{1}{2}\left( e^{i\phi },0,\sqrt{2},0,-e^{-i\phi }\right) ^{T}$ 
 is left invariant
 by the elements in the isotropy group
 \begin{eqnarray}
     H &=&\{\pm I,\pm a,\pm b,\pm c,  \label{c} \\
       &&\pm d,\pm e,\pm f,\pm g,  \nonumber \\
       &&\pm d^2,\pm e^2,\pm f^2,\pm g^2 \}.  \nonumber
 \end{eqnarray}
 The spin rotations $a=u(z,\pi)$, $b=u(y,\pi)u(z,\phi +\pi /2)$ and $c=ba$
 satisfy $a^{2}=b^{2}=c^{2}=-I$, while 
 $d=u(z,\pi/4+\phi/2)u(y,\pi/2)u(z,\pi/4-\phi/2)$,
 $e=-da$, $f=-ad$ and $g=-ada$ satisfy $d^3=e^3=f^3=g^3=-I$. Each
 element in the first, second, third row is associated with an additional
 phase change $2n\pi,2\pi/3+2n\pi,4\pi/3+2n\pi$ respectively.
 It is a discrete group, and $H_{0}$ consists of the identity
 $\left(0,I\right) $ only. The fundamental theorems identify that
 \begin{equation}
     \pi _{1}(M)=H,\qquad \pi _{2}(M)=0\text{, (spin-2 }C\text{ 
 state).}  \label{p2}
 \end{equation}
 The elements in the fundamental group are non-commutative, for example
 $ab=-c\neq ba$. The criterion for the topological equivalence of defects
 applies in the most general case in terms of conjugacy classes of the
 fundamental group. It is thus necessary to classify the group into the
 following conjugacy classes: $\{I \}_n$, $\{-I\}_n$,
 $\{ \pm a,\pm b,\pm c\}_n$, $\{ d,e,f,g\}_{n+1/3}$, $\{-d,-e,-f,-g\}_{n+1/3}$,
 $\{d^2,e^2,f^2,g^2\}_{n+2/3}$, $\{-d^2,-e^2,-f^2,-g^2\}_{n+2/3}$
 with the subscripts standing for the winding numbers of the defects.
 Physically this indicates the feasibility of creating 
 not only vortices with any integer winding number but also fractional quantum 
 vortices. The homology theory \cite{Trebin} assembles the conjugacy classes 
 further into $3$ sets for each $n$, in which the defects are labeled by 
 the winding numbers $n,n+1/3,n+2/3$ respectively \cite{Harri}.

 Interesting features of this non-Abelian fundamental group include the topological
 instability of the defects and their interaction, i.e., entanglement when two
 of them are brought to cross with each other \cite{Mermin}. Two defects in the
 same conjugacy classes can be continuously converted into one another by local
 surgery. In addition, the defects in the same homology set can be 
 transformed into
 each other via a {\it catalyzation} process consisting of splitting a line
 singularity into two and recombining them beyond a third one \cite{Trebin}.

 Like the quaternion group $Q$ for biaxial nematics, the fundamental group 
 (\ref{c}) is the {\it lift} of a point group in $R\times \rm{SU(2)}$. 
 To find the remaining discrete symmetry group for the cyclic state, 
 and, in addition, to clarify the controversial identification of the OPS for spin-1 case,
 in the remaining of this paper we turn to
 describe the system in terms of rotations in $\rm{SO(3)}$, e.g., two elements $
 \pm u(z,\alpha )$ in $\rm{SU(2)}$ are mapped into one $R(z,\alpha )$ in $\rm{SO(3)}$
 with $R(z,\alpha +2\pi )=R(z,\alpha )$.

 {\em Order Parameter Spaces: }The OPS for $F=1$ polar state was identified
 as ${\rm U(1)}\times S^{2}$ in Refs. \cite{Ho,Khawaja}. An extra $Z_{2}$ symmetry
 was claimed in Ref. \cite{Zhou1} so the author concluded the OPS as 
 ${\rm U(1)} \times S^{2}/Z_{2}$. Here we show that previous studies are 
 incorrect. Taking the group $G$ as ${\rm U(1)}_{G}\times {\rm SO(3)}_{S}$ where the 
 subscripts stand for the gauge and spin symmetries respectively,
 we see that the isotropy group $H$ consists of two separate parts, 
 $\left\{ \left(e^{i0},R(z,\alpha) \right) \right\}$ and 
 $\left\{ \left(e^{i\pi},R(y,\pi)R(z,\alpha) \right) \right\}$. The rotations in the
 first part constitute the group ${\rm SO(2)}$, while the elements in the second
 part are just those in the group ${\rm O(2)}$ but not in ${\rm SO(2)}$ with determinants
 $-1$. The combination of these two parts gives the full isotropy group as 
 ${\rm O(2)}$ where both gauge and spin symmetries are involved. The OPS is the
 quotient $G/H=\left( {\rm U(1)}_{G}\times {\rm SO(3)}_{S}\right) /{\rm O(2)}_{G+S}$ and
 here it is not possible to apply the fundamental theorem for $G$ is not any
 more simply connected.

 For the ferromagnetic state the group $H$ may be obtained if one notices
 that the $2\pi $ difference in the rotational angle does not give another
 component as it did in the case of ${\rm SU(2)}$. We have $H=\left\{\left(e^{i\theta
 },R(z,\theta ) \right) \right\}$ which is isomorphic to ${\rm U(1)}_{G+S}.$ This means that
 there is a remaining symmetry ${\rm U(1)}$ in the symmetry broken system. The OPS
 is thus factorized as $\left( {\rm U(1)}_{G}\times {\rm SO(3)}_{S}\right)
 /{\rm U(1)}_{G+S}={\rm SO(3)}_{S+G}$.

 The discrete symmetry group of defects in the spin-2 cyclic state $C$ can be
 shown to be isomorphic to the tetrahedral group $T$. We continue to 
 represent $G$ as ${\rm U(1)}_{G}\times {\rm SO(3)}_{S}$. The isotropy
 group (\ref{c}) is shrunk to a group
 of 12 elements if one understands the rotation in the sense of ${\rm SO(3)}$ (i.e.,
 $a=R(z,\pi )$),
 \begin{equation}
     H=\{I,a,b,c,\varepsilon d,\varepsilon e,\varepsilon f,\varepsilon g,
     \varepsilon ^2 d^2,\varepsilon ^2 e^2, \varepsilon ^2 
 f^2,\varepsilon ^2 g^2\},  \label{cc}
 \end{equation}
 where $\varepsilon =\exp (2\pi i/3)$ comes from the gauge transformation
 and $\varepsilon d$, for instance, is an abbreviation for the element 
 $\left(\varepsilon,d \right)$. Three 2-fold rotational axes are $z$ and
 2 lines in $xy$ plane perpendicular
 to each other (which lie on axes $x$ and $y$ if we choose the arbitrary phase
 $\phi =\pi /2$). The elements $\varepsilon d,\varepsilon 
 e,\varepsilon f,\varepsilon g$
 are four 3-fold axes. The symmetries remaining in the symmetry broken 
 states for biaxial nematics and spin-2 cyclic state are shown in 
 Figure 1. The OPS for state $C$ can be
 identified as $\left( {\rm U(1)}_{G}\times {\rm SO(3)}_{S}\right) /T_{G+S}$.

 \begin{figure}[ht]
 \includegraphics[width=8cm]{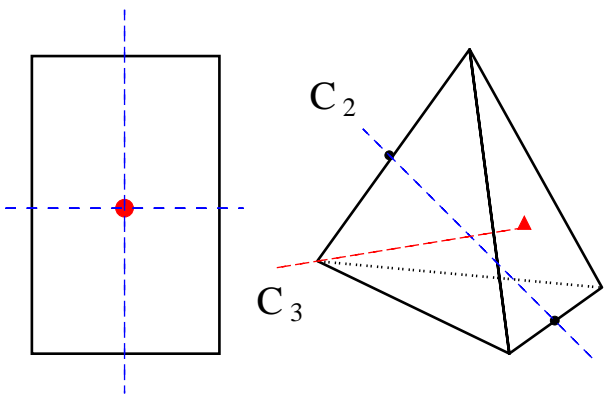}
 \caption{Symmetries of the defects in biaxial nematics ($D_2$) and 
 cyclic state $C$
 in spin-2 condensate($T$). The dot in the center of the rectangle 
 stands for axis $z$.
 The dashed lines represent 2-fold axes, except that with a triangle for
 3-fold axis.}
 \end{figure}

 It should be noted that an applied magnetic field $B$ changes the 
 defect structure severely by reducing  the degenerate family of the 
 spinor. We take again the cyclic state as
 an example. The symmetry group in this case is ${\rm U(1) \times SO(2)}$ because
 the magnetic field chooses its direction automatically as the 
 quantization axis.
  From the spinor $\frac12\left((1+p)e^{i\phi_1},0,\sqrt{2-2p^2}e^{i(\phi_1+
 \phi_2)/2},0,(-1+p)e^{i\phi_2}\right)^T $, where $\phi_{1,2}$ are two arbitrary
 phases and $p \sim B$, we easily see the possibility to create vortices in
 any of the three nonzero components with winding number for $i-$th component
 $n_i$ confined by $n_1+n_5=2n_3$.

 In summary, we have determined the nature of the topological defects 
 in spin-1 and spin-2 condensates. The order parameter spaces are 
 identified as the spaces of the coset of
 the isotropy group $H$ in the transformation group $G$. Topologically 
 stable vortices with
 winding numbers larger than unity may be created in the ferromagnetic 
 state for condensates
 with $F>1$, up to the value $(2F-1)$. The line defects in the spin-2 
 cyclic state $C$ exhibit non-commutative features, resulting e.g. in 
 line defects with winding numbers of $1/3$ and its multiples. 

 The authors acknowledge the Academy of Finland (Project No 50314) for
 financial support. YZ is also supported by NSF of China (Grant Nos. 10175039
 and 90203007). Discussions with O. K\"{a}rki and J. Hietarinta are appreciated.

 \end{document}